\documentclass[conference]{IEEEtran}
\IEEEoverridecommandlockouts
\usepackage[table]{xcolor}
\usepackage{cite}
\usepackage{amsmath,amssymb,amsfonts}
\usepackage{algorithmic}
\usepackage{graphicx}
\usepackage{textcomp}
\usepackage{enumitem}
\usepackage{soul}
\usepackage{longtable}
\usepackage{xspace}
\usepackage[utf8]{inputenc}
\usepackage{booktabs}
\usepackage{tabularx}
\usepackage{float}
\usepackage[htt]{hyphenat}
\usepackage{multirow}
\usepackage{pgfplots}
\usepackage{caption}
\usepackage{tikz}
\usepackage{cleveref}
\pgfplotsset{compat=1.17}

\usepackage{url}
\def\BibTeX{{\rm B\kern-.05em{\sc i\kern-.025em b}\kern-.08em
    T\kern-.1667em\lower.7ex\hbox{E}\kern-.125emX}}
\begin{document}

\newcommand{\ie}{\textit{i.e.,}\xspace}
\newcommand{\eg}{\textit{e.g.,}\xspace}
\newcommand{\etc}{\textit{etc.}\xspace}
\newcommand{\etal}{\textit{et al.}\xspace}

\newcommand{\RQOne}{What usage patterns characterize the use of SonarQube Cloud in the CI/CD pipelines of GitHub projects?\xspace}

\newcommand{\RQTwo}{To what extent SonarQube Cloud projects use default and/or built-in quality gates?\xspace}

\newcommand{\RQThree}{What conditions are verified by the quality gates of SonarQube Cloud projects?\xspace}

\newcommand\giuseppe[1]{\textcolor{red}{#1}}
\newcommand\sabato[1]{\textcolor{black}{#1}}
\newcommand\davide[1]{\textcolor{orange}{#1}}


\title{Dealing with SonarQube Cloud: Initial Results from a Mining Software Repository Study}


\author{\IEEEauthorblockN{Sabato Nocera}
\IEEEauthorblockA{
\textit{University of Salerno}\\
Fisciano, Italy \\
snocera@unisa.it}
\and
\IEEEauthorblockN{Davide Fucci}
\IEEEauthorblockA{
\textit{Blekinge Institute of Technology}\\
Karlskrona, Sweden \\
davide.fucci@bth.se}
\and
\IEEEauthorblockN{Giuseppe Scanniello}
\IEEEauthorblockA{
\textit{University of Salerno}\\
Fisciano, Italy \\
gscanniello@unisa.it}
}

\newcommand\copyrighttext{%
  \footnotesize 979-8-3315-9147-2/25/\$31.00~\copyright2025 IEEE \hfill}

\newcommand\copyrightnotice{%
  \begin{tikzpicture}[remember picture,overlay]
    \node[anchor=south,yshift=40pt] at (current page.south) 
      {\parbox{\dimexpr\textwidth\relax}{\copyrighttext}};
  \end{tikzpicture}%
}


\maketitle


\begin{abstract}
\textit{Background:} Static Code Analysis (SCA) tools are widely adopted to enforce code quality standards. However, little is known about how open-source projects use and customize these~tools. \\
\textit{Aims:} This paper investigates how GitHub projects use and customize a popular SCA tool, namely SonarQube Cloud.\\
\textit{Method:} We conducted a mining study of GitHub projects that are linked through GitHub Actions to SonarQube Cloud projects. \\
\textit{Results:} Among 321 GitHub projects using SonarQube Cloud, 81\% of them are correctly connected to SonarQube Cloud projects, while others exhibit misconfigurations or restricted access. Among 265 accessible SonarQube Cloud projects, 75\% use the organization's default \textit{quality gate}, \ie a set of conditions that deployed source code must meet to pass automated checks. 
While 55\% of the projects use the built-in quality gate provided by SonarQube Cloud, 45\% of them customize their quality gate with different conditions.
Overall, the most common quality conditions align with SonarQube Cloud’s ``Clean as You Code'' principle and enforce security, maintainability, reliability, coverage, and a few duplicates on newly added or modified source code.\\
\textit{Conclusions:} Many projects rely on predefined configurations, yet a significant portion customize their configurations to meet specific quality goals. Building on our initial results, we envision a future research agenda linking quality gate configurations to actual software outcomes (\eg improvement of software security). This would enable evidence-based recommendations for configuring SCA tools like SonarQube Cloud in various contexts. 
\end{abstract}

\copyrightnotice

\section{Introduction}
Automation is a pillar of modern Software Engineering (SE) practices, such as DevOps and DevSecOps~\cite{ozkaya2019devops}.
Particularly, the field of software security identifies a spectrum of automation approaches, from fixed policies (\ie explicitly embedded in a tool/system, without the possibility of changes), to customizable (\ie allowing custom configurations), to dynamic ones (\ie the most flexible that adapt at runtime)~\cite{Edwards08}.
The impact of different automation policies has been extensively studied in the context of end-user security (\eg to suggest policies for the Windows operating system’s automatic updates~\cite{Morris2019}).
However, investigating automation policies---such as the ones embedded in Static Code Analysis (SCA) tool configurations---is also fundamental in SE for improving developers’ experience~\cite{Noda2023}.

SCA tools have become integral to modern software development, particularly in Continuous Integration/Continuous Delivery (CI/CD) pipelines, where they help enforce coding standards and detect quality issues (including security issues) early. Among these tools, SonarQube Cloud~\cite{SonarCloud:2025} stands out as a popular and widely used Software-as-a-Service (SaaS) platform that provides automated code reviews and quality assessments across multiple programming languages~\cite{digkas2018developers,Nocera:2023:Seaa:Sm,Nocera:2024:IcseSeet,baldassarre2020diffuseness,nocera2025software,Nocera:2024:Quatic}.

A central feature of the automation provided by SonarQube Cloud is the \textit{quality gate}~\cite{QualityGate:2025}, which is a configurable set of conditions that code must satisfy to be considered acceptable. These conditions typically include thresholds for~code coverage, code duplication, maintainability, reliability, and~security ratings. SonarQube Cloud tool also offers a (predefined) built-in quality gate, known as the ``Sonar way,'' which enforces best practices on added or modified source code, following the ``Clean as You Code'' principle~\cite{CleanAsYouCode:2025}. Nonetheless, project administrators can also define (customized) non-built-in quality gates to better reflect the quality goals of their projects.

Despite the growing adoption of SonarQube Cloud in open-source development (\eg it is used by more than 22k GitHub users~\cite{sonarqube-scan-action}), there is a limited empirical understanding of how developers configure and use it in practice~\cite{Bennett2024, vassallo2020developers}. Questions remain on the extent to which predefined configurations are retained, how quality gates are customized, and which quality conditions are prioritized across projects. Answering these questions (\ie understanding how developers interact with SonarQube Cloud) is important, for example, to bridge the gap between tool design and practice, guide tool vendors in refining predefined configurations, and improve automation.

In this paper, we provide \textit{initial results} from a mining study of 321 GitHub projects using SonarQube Cloud with the final goal of identifying challenges and opportunities in SE research and practice. Among other things, our results reveal a nuanced landscape: while many projects rely on predefined configurations (\eg organization's default quality gate or SonarQube Cloud's built-in quality gate), a significant portion use customized quality gates, reflecting diverse quality assurance strategies. These insights contribute to a better understanding of how SCA tools are operationalized.
Our study lays the groundwork for evidence-based recommendations for configuring SCA tools and future research linking configuration choices to software outcomes (\eg improvement of software quality characteristics such as security).  




    
    
    


\section{Background and Related Work} \label{sec:StateOfArt} 
In this section, we provide background on SonarQube Cloud and related work investigating SCA tool configurations.

\subsection{SonarQube Cloud 
} \label{sec:SonarCloud}

SonarQube Cloud (formerly known as SonarCloud) is a SaaS automated code review and SCA tool designed to detect coding issues in many programming languages~\cite{SonarCloud:2025}. It is conceived to be used in {CI/CD} pipelines, and it integrates with all leading CI/CD systems (including GitHub Actions). To enable SonarQube Cloud in the CI/CD pipeline of a software project, it is necessary to install and configure a component called \textit{scanner} and connect that software project with a SonarQube Cloud project hosted on the SonarQube Cloud platform~\cite{SonarCloudExplore:2025}.

A GitHub project can connect to a SonarQube Cloud project by using one of the following GitHub Actions: \texttt{sonarcloud-github-action}~\cite{sonarcloud-github-action} or \texttt{sonarqube-scan-action}~\cite{sonarqube-scan-action}.
\footnote{Although deprecated and planned for removal,  \texttt{sonarcloud-github-action} is still used alongside its replacement, \ie \texttt{sonarqube-scan-action}, in GitHub projects.} 
To enable the connection, the GitHub project must specify the \textit{keys} of the SonarQube Cloud project and its organization. These keys are specified via the \texttt{sonar.projectKey} and \texttt{sonar.organization} attributes, which can be provided either in a \texttt{sonar-project.properties} file or within the configuration of the GitHub Action. For instance, the GitHub project \texttt{\url{https://github.com/apache/cloudberry}} specified the SonarQube Cloud keys in its \texttt{sonar-project.properties} file: \texttt{sonar.projectKey=apache\_cloudberry} and \texttt{sonar.organization=apache}.

Once SonarQube Cloud is configured into a CI/CD pipeline, upon the occurrence of specified events (\eg push or merge request), the scanner analyzes the application's source code against a set of conditions called \textit{quality gate}~\cite{QualityGate:2025}. These conditions may include thresholds for code coverage, code duplication, maintainability, reliability, and security ratings. Maintainability issues (or \textit{code smells}) indicate patterns that make the source code confusing or difficult to maintain. Reliability issues (or \textit{bugs}) are coding errors likely to cause failures at runtime. Security-related issues include both \textit{vulnerabilities}, requiring immediate remediation, and \textit{hotspots}, highlighting potentially risky code that should be reviewed. Each coding issue is assigned a severity level, with levels ranging from least to most severe: \textit{info}, \textit{minor}, \textit{major}, \textit{critical}, and \textit{blocker}.  

SonarQube Cloud shows a \textit{Passed} status when the analysis meets or exceeds the quality gate conditions; otherwise, it shows a \textit{Failed} status. Each SonarQube Cloud project has a single quality gate definition that is activated at any given time~\cite{SonarStandards:2025}. 
Whenever a SonarQube Cloud project is created, its quality gate is set to the \textit{default}. Project administrators choose the default quality gate for their projects among those available. SonarQube Cloud provides for every project its own \textit{built-in} quality gate, \ie ``Sonar way.'' According to SonarQube Cloud documentation~\cite{SonarCloud:2025}, the built-in quality gate is suitable for most projects; nonetheless, project administrators can also define and choose a \textit{non-built-in} quality gate. 

\subsection{Empirical Studies on Static Code Analyzers configurations} 
A survey of software developers~\cite{Bennett2024} (n = 1,263) using Static Application Security Testing tools shows that, despite their popularity, 54\% do not configure them, using out-of-the-box rule sets.
Among the developers who change predefined configurations, the vast majority enable or disable predefined rules rather than create new ones.
These changes to predefined configurations are associated with prolonged experience with the tools.
Interestingly, SonarQube was the most popular tool in the study, as reported by 59\% of the participants. 

Zampetti et al.~\cite{Zampetti2017} studied SCA and their configurations mined from CI/CD pipelines of 20 GitHub projects.
From the projects' commit history, the authors extracted changes made to configuration files and extracted categories of rules activated or deactivated at least once.
Their results show that many SCA tools are configured to break the build, while a small minority only raise warnings without interrupting the pipeline.
In 17 projects, the configurations are customized by activating additional rules beyond the predefined ones; however, there is a variation, between 40 and 50\%, of activated rules across tools.
For five projects, the configurations are never changed after the first time, whereas for projects with continued changes over time, rules are never disabled or removed.
From these results, developers’ motivation to change configurations is to limit additional build-breaking checks that are not fully understood.

Besides affecting build pipelines, different SCA configurations can also impact the overall security of a software product.
Piskachev et al.~\cite{Piskachev2023} conducted a user study involving 40 practitioners with varying roles, including developers and security experts, spanning from 3 to 10+ years of experience.
The study compared the performance of participants in addressing known vulnerabilities in a software-under-test using the predefined SCA configuration vis-à-vis allowing them to customize the tool.
The results show that participants who changed the configuration were able to address 76\% of vulnerabilities---a 15\% improvement over the predefined configuration.
Furthermore, the authors collected qualitative evidence through post-task interviews to understand the strategies employed by participants to modify the tool configuration.
The findings revealed that participants adopted an iterative approach---they activate a different subset of rules at each iteration and study their output to address possible vulnerabilities.

SCA tools configuration is reported as one of the main pain points, together with false negatives, in the results of a semi-structured interview with 20 practitioners from diverse backgrounds (security, product management, 
development)~\cite{ami2024false}. 
Conversely, participants expressed ease of configuration as one of the most important characteristics of an SCA tool. 

Our contribution is transversal to the research introduced above. We provide initial evidence on how developers use a popular SCA tool (\ie SonarQube Cloud) in practice, and, based on it, we identify opportunities in SE research and practice to support developers when dealing with such tools. 

\section{Study Design 
} \label{sec:design}

In this section, we present the design of our study.

\subsection{Goal and Research Question}

By leveraging the Goal-Question-Metric (GQM) method~\cite{basili1992software}, we formalize the goal of our study as~follows:

\begin{itemize}
    \item[] \textbf{Analyze} the use of SonarQube Cloud \textbf{for the purpose of} \textit{(i)} understanding usage patterns, \textit{(ii)} assessing the degree of customization of quality gates, and  \textit{(iii)} identifying the conditions verified by such quality gates 
    \textbf{from the point of view of} practitioners and researchers \textbf{in the context of} open-source projects publicly hosted on GitHub that use SonarQube Cloud as part of their CI/CD pipelines.
\end{itemize}
Based on the above-mentioned goal, we formulated the following three Research Questions (RQs):
\begin{itemize}[leftmargin=1cm]
    \item[\textbf{RQ1.}] \textit{\RQOne} 
\end{itemize}
This RQ aims to understand the usage patterns characterizing the use of SonarQube Cloud in the CI/CD pipelines of GitHub projects. \sabato{
We derived such patterns while exploring} 
how GitHub projects connect to the projects hosted~on SonarQube Cloud and \sabato{incur} eventual misconfigurations or errors.
\begin{itemize}[leftmargin=1cm]
    \item[\textbf{RQ2.}] \textit{\RQTwo}
\end{itemize}
This RQ aims to assess the extent to which SonarQube Cloud projects use the organization's default quality gate and SonarQube Cloud's built-in quality gate, as well as the proportion of projects that do not use them. In other words, this RQ investigates the degree of customization in the quality gates used by SonarQube Cloud projects.

\begin{itemize}[leftmargin=1cm]
    \item[\textbf{RQ3.}] \textit{\RQThree}
\end{itemize}
This RQ aims to identify the quality conditions defined and verified by the quality gates of SonarQube Cloud projects. This allows for investigating which quality characteristics (\eg maintainability, reliability, or security) developers choose to prioritize in CI/CD pipelines.

\subsection{Study Context and Planning}

The context of our study consists of open-source software projects publicly hosted on GitHub and using SonarQube Cloud in their CI/CD pipelines to enforce code quality standards. We considered GitHub because it is currently the most popular version-control and software-development-hosting platform for personal and professional use~\cite{StackOverflow:2022}.  

To search for software projects using SonarQube Cloud, we leveraged GitHub's dependency graph feature~\cite{GitHubDependencyGraph:2025}. It allows retrieving all the dependencies and dependents of a given software repository, \ie the repositories a given repository depends on and the repositories that depend on it, respectively. The information about the dependencies and dependents of a repository is automatically inferred from the manifest and lock files, \ie files specifying project dependencies such as \texttt{.yaml} for GitHub Actions. In our case, we leveraged GitHub's dependency graph to retrieve all GitHub repositories dependent on \texttt{sonarcloud-github-action}~\cite{sonarcloud-github-action} or \texttt{sonarqube-scan-action}~\cite{sonarqube-scan-action}.  We retrieved 34,975 
GitHub repositories dependent on \texttt{sonarcloud-github-action} and 12,986 dependent on \texttt{sonarqube-scan-action}. 


The analysis of data retrieved from GitHub can lead to conclusions not representative of the open-source projects  
intended to be investigated~\cite{Kalliamvakou:2014}. To deal with such an issue, we selected those repositories that met the following criteria:
\begin{enumerate}
    \item \textit{Not archived and with at least one commit made in the month preceding the query date (16 April 2025).} This was to avoid selecting software projects that were inactive~\cite{Kalliamvakou:2014}.
    \item \textit{With at least 100 stars, at least 100 commits, and at least one fork.} This was to mitigate the risk of selecting personal projects~\cite{Kalliamvakou:2014}. We opted for such thresholds as they are common in mining software repository studies (\eg~\cite{Nocera:2023:Seaa:Stream,ibrahim2021study,alfadel2023empirical}).   
    \item \textit{Not fork.} This was to limit the risk of selecting duplicate projects~\cite{Kalliamvakou:2014}. Forks inherit most of their data from the parent repository~\cite{hadian2022exploring} and are likely inactive~\cite{stuanciulescu2015forked}.
\end{enumerate}
We ended up with 231 repositories dependent on \texttt{sonarcloud-github-action} and 149 dependent on \texttt{sonarqube-scan-action}. 
Since a repository could depend on both \texttt{sonarcloud-github-action} and \texttt{sonarqube-scan-action}, we discarded 59 duplicates among the GitHub repositories retrieved with GitHub's dependency graph. Eventually, we analyzed 321 unique GitHub projects using SonarQube Cloud through \texttt{sonarcloud-github-action} and/or \texttt{sonarqube-scan-action}.

\subsection{Data Analysis}

This section presents the data analysis arranged by RQ.

\noindent \textbf{RQ1.} We performed the following steps for each project:

\begin{enumerate}
    \item We examined the content of the \texttt{sonar-project.properties} file and GitHub Actions files to identify the keys required for connecting the GitHub project with the SonarQube Cloud project.
    \item We manually inspected the content of the repository to observe any pattern in the usage of SonarQube Cloud. We searched SonarQube Cloud references through the GitHub Code Search feature available for every repository; we used ``\texttt{sonar}'' as a search string and examined the search results (\eg files in which that string was found). If we did not find search results this way, we also examined the output of GitHub Actions workflows to gain any possible insights on the usage of SonarQube Cloud.
    Among other things, this allowed us to check and, eventually, correct the SonarQube Cloud keys that had been previously retrieved. We collected information about usage patterns through \textit{open-coding}~\cite{miles1994qualitative}. \sabato{The first author conducted it and iteratively refined the resulting codes.}
    \item We verified whether the SonarQube Cloud project connected to the GitHub project could be queried through the SonarQube Cloud API~\cite{SonarCloudApi:2025}. This was to understand if the identified SonarQube Cloud project allowed users outside its organization to access information about its code standards, including quality gates.
\end{enumerate}

\noindent \textbf{RQ2.} We exploited the SonarQube Cloud API~\cite{SonarCloudApi:2025} to collect data on the quality gates used by the identified and accessible SonarQube Cloud projects. For each project, we determined whether the associated quality gate was the default one and whether it was built-in. We then computed the number and percentage of projects using quality gates that were default or non-default, built-in or non-built-in. Specifically, we analyzed all combinations of these attributes, distinguishing between quality gates that were: \textit{(i)}~both default and built-in, \textit{(ii)}~default but not built-in, \textit{(iii)}~built-in but not default, and \textit{(iv)}~neither default nor built-in.

\noindent \textbf{RQ3.} We exploited the SonarQube Cloud API~\cite{SonarCloudApi:2025} to collect data on the conditions verified by the quality gates of the identified SonarQube Cloud projects. We categorized each condition based on the classification provided by SonarQube Cloud documentation (\eg complexity, coverage, security)~\cite{SonarCloudMeasures:2025}. For each condition and each category, we counted 
the number and percentage of projects whose quality gates included it. 

\section{Results 
} \label{sec:Results} 

Below, we report the results by RQ.

\subsection{RQ1. \RQOne}


\begin{table}[t]
\centering
\caption{Usage patterns of SonarQube Cloud.}
\label{tab:rqone}
\rowcolors{2}{gray!10}{white}
\setlength{\tabcolsep}{5pt}
\scriptsize
\begin{tabular}{|l|c|}
\hline
\rowcolor{gray!25}
\textbf{Usage Pattern} & \textbf{Count (\%)} \\
\hline
SonarQube Cloud connected correctly & 260 (81\%) \\
SonarQube Cloud project absent or private & 47 (14.6\%) \\
SonarQube Cloud keys retrieved from GitHub Actions workflows & 26 (8.1\%) \\
GitHub project connected to several SonarQube Cloud projects & 17 (5.3\%) \\
SonarQube Cloud keys defined through environment variables & 17 (5.3\%) \\
SonarQube Cloud keys not found & 12 (3.7\%) \\
Integration of SonarQube Cloud with SonarLint & 11 (3.4\%) \\
SonarQube Cloud keys retrieved through GitHub Code Search & 8 (2.5\%) \\
Integration of SonarQube Cloud with Gradle & 7 (2.2\%) \\
SonarQube Cloud keys defined incorrectly in files & 6 (1.9\%) \\
Integration of SonarQube Cloud with Maven & 4 (1.2\%) \\
SonarQube Server used in place of SonarQube Cloud & 3 (0.9\%) \\
SonarQube Cloud project not configured & 1 (0.3\%) \\
SonarQube Cloud keys commented out & 1 (0.3\%) \\
\hline
\end{tabular}
\end{table}

In Table~\ref{tab:rqone}, we report the identified usage patterns of SonarQube Cloud by GitHub projects. In the first place, 81\% of the GitHub projects were connected correctly to SonarQube Cloud. In the other projects, we acknowledged the following problems with the connection to SonarQube Cloud:
\begin{itemize}
    \item The SonarQube Cloud project was absent or made private in 14.6\% of the GitHub projects. SonarQube Cloud does not explain if a given project is inaccessible because it does not exist or the user is not granted access permissions. 
    \item We could not find SonarQube Cloud keys in 3.7\% of the GitHub projects. We found no reference to SonarQube Cloud through either GitHub Code Search or GitHub Actions.
    \item The SonarQube Cloud project was not correctly configured in 0.3\% of the GitHub projects, and, in another 0.3\% of the GitHub projects, the SonarQube Cloud keys were commented out in the GitHub Action file (meaning that the GitHub Action could not be executed anyway).
\end{itemize}

During the manual inspection of the repositories, we retrieved some SonarQube Cloud keys from sources other than the \texttt{sonar-project.properties} file or GitHub Actions files. We retrieved SonarQube Cloud keys from the output of GitHub Actions workflows in 8.1\% of the GitHub projects and through GitHub Code Search in 2.5\% of them. We also acknowledged that 5.3\% of the GitHub repositories defined SonarQube Cloud keys through environment variables (\eg \texttt{\$\{\{env.sonar-project-key\}\}}). We also observed that, in 1.9\% of GitHub repositories, SonarQube Cloud keys were defined incorrectly, meaning that they could not refer to an existing SonarQube Cloud project.

Some GitHub projects integrated SonarQube Cloud with other technologies. In detail, 2.2\% of GitHub projects integrated SonarQube Cloud with the \textit{Gradle} package manager and 1.2\% with the \textit{Maven} package manager; the former declared the SonarQube Cloud keys in the \texttt{pom.xml} files, while the latter declared them in the \texttt{build.gradle} files. Also, 3.4\% of GitHub projects connected SonarQube Cloud with \textit{SonarLint}~\cite{SonarLint:2025}. The latter is a linter, also known as \textit{SonarQube for IDE}, that can be plugged into IDEs and is developed by the same organization as SonarQube Cloud, namely SonarSource~\cite{SonarSource:2025}. SonarLint can detect a subset of SonarQube Cloud issues while developers code, helping them address such issues before even committing the code. Connecting SonarLint with SonarQube Cloud allows (parts of) the configuration of the two tools to be shared.
There is also another version of SonarQube Cloud, namely \textit{SonarQube Server}~\cite{SonarQubeServer:2025}, which is used \textit{on-premises}, \ie it is installed and managed by an organization's own physical hardware. We identified the use of SonarQube Server in place of SonarQube Cloud in 0.9\% of the GitHub projects.

We found that 5.3\% of GitHub projects were connected and using more than one SonarQube Cloud project. For example, the GitHub project \texttt{\url{https://github.com/WE-Kaito/digimon-tcg-simulator}} was connected to two SonarQube Cloud projects: \texttt{\url{https://sonarcloud.io/summary/overall?id=we-kaito_digimon-tcg-simulator-backend}} and \texttt{\url{https://sonarcloud.io/summary/overall?id=we-kaito_digimon-tcg-simulator-frontend}}; the former seems to be related to the back-end of the project, while the latter to its front-end, suggesting that different components of the same software project may require different code standards. 

\sabato{As a result, we found that 321 GitHub projects were correctly connected to 328 SonarQube Cloud projects. This means that there were some GitHub projects connected to more than one SonarQube Cloud project. Among the SonarQube Cloud projects,}
63 could not be queried through the SonarQube Cloud API due to a lack of permission (\ie we were not part of the organization owning those SonarQube Cloud projects). Those 63 SonarQube Cloud projects were connected to 30 GitHub projects, owned by 21 different GitHub users. We attempted to find reasons why these SonarQube Cloud projects restricted access through the SonarQube Cloud API. We speculate that, in most cases, this might be because the project operates in a security-sensitive application domain (\eg e-commerce or cryptocurrency) and, therefore, a willingness to limit available information on such projects. For example, five of these GitHub projects were owned by \textit{bitwarden}~\cite{bitwarden}, an organization providing security solutions such as password and secret managers. 
Eventually, we could retrieve the information about the code standards, including quality gates, of 265 SonarQube Cloud projects, which were connected to a total of 230 GitHub projects.

\subsection{RQ2. \RQTwo}


\begin{table}[t]
\centering
\scriptsize
\caption{
SonarQube Cloud projects using and not using default and/or built-in quality gates.}
\label{tab:rqtwo}
\rowcolors{2}{gray!10}{white}
\begin{tabular}{|ll|r|r|r|}
\hline
\rowcolor{gray!25}
\multicolumn{2}{|c|}{\textbf{Quality Gate}} & \textbf{Default: True} & \textbf{Default: False} & \multicolumn{1}{c|}{\textbf{\textit{Total}}} \\
\hline

  & \textbf{True}  & 141 (53\%) & 4 (2\%)   & \textit{145 (55\%)} \\
\multirow{-2}{*}{\textbf{Built-in:}}   & \textbf{False} & 57 (22\%)  & 63 (24\%) & \textit{120 (45\%)} \\
\hline
\rowcolor{gray!15}
\multicolumn{2}{|c|}{\textbf{\textit{Total}}} & \textit{198 (75\%)} & \textit{67 (25\%)} & \textit{265 (100\%)} \\
\hline
\end{tabular}
\end{table}

We analyzed the use of default and/or built-in quality gates among the 265 open-source projects using SonarQube Cloud (see Table~\ref{tab:rqtwo}). Our analysis reveals that 141 projects (53\%) used a quality gate that was both default and built-in. Conversely, 63 projects (24\%) applied a quality gate that was neither default nor built-in, reflecting a fully customized configuration for those projects. Other 57 projects (22\%) used the default quality gate but with a non-built-in version, while only 4 projects (2\%) applied the built-in quality gate without being set as the default. In total, 198 projects (75\%) relied on the default quality gate, and 145 (55\%) used the built-in one. These findings suggest that while most projects rely on default and built-in configurations---likely due to their convenience and general applicability---a non-negligible number of projects customize those configurations. In particular, 67 projects (25\%) did not retain the default quality gate of their organization, and 120 projects (45\%) followed a customized configuration for the quality gate.

\subsection{RQ3. \RQThree}

\begin{table}[t]
\centering
\caption{Identified quality conditions 
(in bold those in the built-in quality gate).}
\label{tab:criteria}
\rowcolors{2}{gray!10}{white}
\setlength{\tabcolsep}{5pt}
\scriptsize
\begin{tabular}{|l|l|c|}
\hline
\rowcolor{gray!25}
\multicolumn{1}{|c|}{\textbf{Category}} & \multicolumn{1}{c|}{\textbf{Condition}} & \multicolumn{1}{c|}{\textbf{Count (\%)}} \\
\hline
Complexity & Cognitive complexity & 2 (0.75\%) \\
\hline
Coverage & Condition coverage on new code & 12 (4.53\%) \\
 & \textbf{Coverage on new code} & \textbf{205 (77.36\%)} \\
 & Coverage & 25 (9.43\%) \\
 & Unit test success density & 1 (0.38\%) \\
 & Uncovered lines on new code & 1 (0.38\%) \\
\hline
Duplications & \textbf{Duplicated lines density on new code} & \textbf{234 (88.30\%)} \\
 & Duplicated lines density & 18 (6.79\%) \\
 & Duplicated block on new code & 1 (0.38\%) \\
 & Duplicated lines on new code & 1 (0.38\%) \\
\hline
Issues & Issues on new code & 2 (0.75\%) \\
 & Blocker issues on new code & 8 (3.02\%) \\
 & Blocker issues & 4 (1.51\%) \\
 & Critical issues on new code & 4 (1.51\%) \\
 & Critical issues & 6 (2.26\%) \\
 & Major issues on new code & 2 (0.75\%) \\
 & Minor issues on new code & 1 (0.38\%) \\
\hline
Maintainability & \textbf{Maintainability rating on new code} & \textbf{236 (89.06\%)} \\
 & Maintainability rating & 16 (6.04\%) \\
 & Code smells & 1 (0.38\%) \\
 & Code smells on new code & 8 (3.02\%) \\
\hline
Reliability & \textbf{Reliability rating on new code} & \textbf{235 (88.68\%)} \\
 & Reliability rating & 18 (6.79\%) \\
 & Bugs & 7 (2.64\%) \\
 & Bugs on new code & 10 (3.77\%) \\
\hline
Security & \textbf{Security rating on new code} & \textbf{241 (90.94\%)} \\
 & Security rating & 21 (7.92\%) \\
 & Vulnerabilities & 7 (2.64\%) \\
 & Vulnerabilities on new code & 5 (1.89\%) \\
\hline
Security Review & Security review rating on new code & 4 (1.51\%) \\
 & Security review rating & 3 (1.13\%) \\
 & Security hotspots reviewed & 14 (5.28\%) \\
 & \textbf{New security hotspots reviewed} & \textbf{232 (87.55\%)} \\
\hline
\end{tabular}
\end{table}

We found a total of 33 different quality conditions being verified by the quality gates of SonarQube Cloud projects. We report these conditions in Table~\ref{tab:criteria}, along with the number (and percentage) of projects adopting each, grouped by the category indicated by SonarQube Cloud~\cite{SonarCloudMeasures:2025}.  
We have to mention that a given condition could be verified differently across projects by applying different thresholds. For example, one SonarQube Cloud project may consider the condition \textit{vulnerabilities} met only when the number of vulnerabilities is zero, whereas another may allow up to one vulnerability.

The most commonly adopted conditions are those included in the built-in quality gate (see bold entries in  Table~\ref{tab:criteria}) and are applied to \textit{new code}. In fact, SonarQube Cloud enforces the \textit{Clean as You Code} practice, which is based on the principle that new code (\ie code that was added or modified recently) needs to comply with quality standards~\cite{CleanAsYouCode:2025}. Since adding new code usually involves changes in existing code, analyzing and cleaning new code enables the gradual improvement of the overall quality of the codebase. 

Among the conditions of the built-in quality gate, \textit{security rating on new code} is the most prevalent, enforced in 90.94\% of projects. It assesses newly introduced vulnerabilities on a five-level scale from A (no vulnerabilities) to E (at least one blocker vulnerability). Next, \textit{maintainability rating on new code} appears in 89.06\% of quality gates. It assigns a grade from A to E by comparing the estimated effort for fixing the detected maintainability issues (technical debt) with the effort needed to develop the same code. \textit{Reliability rating on new code}, included in 88.68\% of quality gates, similarly grades newly introduced bugs from A (no bugs) to E (at least one blocker bug). In 88.30\% of quality gates, \textit{duplicated lines density on new code} was applied to verify the proportion of duplicate lines within the newly added or modified code. \textit{New security hotspots reviewed}, enforced in 87.55\% of projects, verifies the percentage of reviewed security hotspots on new code. Finally, we found \textit{coverage on new code} in 77.36\% of quality gates. It combines line and condition coverage to evaluate the extent to which new or updated code is covered. 

The conditions included in the built-in quality gate are also widely adopted in non-built-in quality gates. Considering that 45\% of SonarQube Cloud projects use non-built-in quality gates, the least adopted built-in condition (\textit{coverage on new code}) was enforced by 77.36\% of projects, while the most adopted one (\textit{security rating on new code}) was enforced by 90.94\% of projects. This difference may also reflect the prioritization of certain quality characteristics over others (\eg security over code coverage).

Beyond the built-in conditions, Table~\ref{tab:criteria} shows various other quality conditions used across projects. However, these are adopted with much lower frequency, \ie each by less than 10\% of projects. This suggests that, although some projects may use quality gates with custom conditions, they tend to rely primarily on those conditions enforced by SonarQube Cloud.

\section{Discussion}\label{sec:Discussion}
This section discusses the results, highlighting their implications and presenting potential threats to validity.

\subsection{Implications of the Results}
One of the main results of our study indicates that most practitioners do not change the predefined configuration of SonarQube Cloud, which is coherent with past research~\cite{Bennett2024,Zampetti2017}. On the other hand, a significant portion of the analyzed projects customize SonarQube Cloud configurations. This seems to suggest that predefined SCA tool configurations might not always be suitable and projects might require \textit{dynamic configurations}~\cite{Edwards08}. 
To better understand this phenomenon, researchers should investigate the needs and motivation for deviating from predefined configurations (\eg through qualitative studies).
Similarly, approaches based on machine learning and artificial intelligence could be leveraged to create configurations based on such needs dynamically. 

Our results also encourage researchers to investigate SCA tool customization patterns. For example, different contexts could drive the prioritization of different parts of a configuration (\eg enforcing quality conditions related to security rather than maintainability).
Researchers could also be interested in developing taxonomies capturing different customization patterns. This would support the improvement of predefined SCA tool configurations, as well as their customization.

Despite existing research showing that integrating several SCA tools improves software outcomes (\eg vulnerability detection~\cite{GOSEVAPOPSTOJANOVA201518} or software bill of material generation~\cite{Nocera:2024:Icsme}), we found little evidence of SonarQube Cloud integrations with other tools, such as SonarLint.
We foster research to investigate the issues arising when integrating several SCA tools with respect to their configurations (\eg conflicting rules or conditions).
Accordingly, a further avenue of research is devising dynamic configurations tailored for the combined use of multiple SCA tools, rather than a single SCA tool.


\subsection{Threats to Validity}

We acknowledge some threats to the validity of our results. 
In the following, we discuss these threats according to the guidelines by Wohlin~\etal~\cite{Wohlin:2012}. We would like to mention that no threats to internal validity are presented because we 
did not investigate causal relationships. 

\textbf{Construct Validity.} Our identification of SonarQube Cloud usage relies on GitHub’s dependency graph and manual inspection of configuration files. Similarly to past studies (\eg~\cite{Nocera:Icsme:2023,zhao2024empirical}), we used GitHub's dependency graph to retrieve repositories, in our case dependent on \texttt{sonarcloud-github-action} and/or \texttt{sonarqube-scan-action}.
GitHub's dependency graph is prone to inaccuracies that can affect the results~\cite{bifolco:2024,bifolco2025empirical}. 

\textbf{Conclusion Validity.} Our analysis's correctness depends on the SonarCloud API's accuracy and the interpretation of quality gate configurations. We mitigated this possible threat by cross-validating data sources and manually inspecting results and repositories.

\textbf{External Validity.} We studied open-source GitHub projects meeting specific activity and popularity thresholds. As such, our findings may not be generalized to private or enterprise projects or to GitHub projects with different characteristics.


\section{Remarks and Vision for Future Research}\label{sec:conclusion}
This paper presents initial results from a mining software repository study on how open-source developers use and configure a popular Static Code Analysis (SCA) tool, namely SonarQube Cloud. By analyzing 321 GitHub projects, we found that, while most projects rely on predefined SonarQube Cloud configurations (\ie default and built-in quality gates), a significant portion of those projects customize their configurations to better align with specific quality goals. The most commonly enforced conditions---such as security, maintainability, and reliability ratings on new code---reflect a strong adherence to SonarQube’s ``Clean as You Code'' principle.

Our findings highlight that, while predefined configurations offer convenience and broad applicability, many projects require customized SCA tool configurations to meet their unique needs. This suggests that many developers are actively shaping how SCA tools like SonarQube Cloud are used in their development workflows. 

As a vision for future work, we propose investigating the effectiveness of different quality gate configurations by linking them to downstream software outcomes, such as the improvement of software quality characteristics like security. This investigation could provide actionable insights into which configurations yield the greatest benefits in practice, ultimately guiding tool developers and practitioners toward more evidence-based quality assurance strategies.

\section*{Data Availability}

The 
data are available online~\cite{ReplicationPackage}.

\section*{Acknowledgment}
This project has been financially supported by the European Union NEXTGenerationEU project and by the Italian Ministry of the University and Research MUR, a Research Projects of Significant National Interest (PRIN) 2022 PNRR, project n. D53D23017310001 entitled ``Mining Software Repositories for enhanced Software Bills of Materials'' (MSR4SBOM), and the SEcure software engineering through Sensible AutoMation (SESAM) founded by KKS.

\bibliographystyle{IEEEtran}
\bibliography{bibliography}

\vspace{12pt}

\end{document}